\documentstyle[psfig,nicV]{article}
\begin{document}

\newcommand{\ned}{$^{22}$Ne}
\newcommand{\nedb}{$^{22}$Ne~}
\newcommand{\msb}{$M_{\odot}$~}
\newcommand{\ms}{$M_{\odot}$}
\newcommand{\cd}{$^{12}$C}
\newcommand{\cdb}{$^{12}$C~}
\newcommand{\ct}{$^{13}$C}
\newcommand{\ctb}{$^{13}$C~}
\newcommand{\s}{$s$}
\newcommand{\p}{$p$}
\newcommand{\rr}{$r$}
%
%Begin Heading
%

\heading{%
A New Astrophysical Interpretation of the Si and Ti Isotopic
Compositions of Mainstream SiC Grains from Primitive Meteorites
%
%End Heading
}
\par\medskip\noindent
%
%Begin Author names
\author{%
Maria Lugaro$^{1}$, Roberto Gallino$^{2}$, Maurizio Busso$^{3}$,
Sachiko Amari$^{4}$, Ernst Zinner$^{4}$

%End Author names
}
%First address
\address{
Department of Mathematics, Monash
University, Clayton 3168, Victoria, Australia
}
% Second Address
\address{
Dipartimento di Fisica Generale,
Universita' di Torino, Via P. Giuria 1, I-10125 Torino, Italy
}
% Third Address
\address{
Osservatorio Astronomico di Torino, I-10025 Pino Torinese,
Italy
}
% Fourth Address
\address{
McDonnell Center for the Space Sciences and Physics
Department, Washington University, St Louis, MO 63130, USA
}
\begin{abstract}
Mainstream presolar SiC grains from primitive meteorites show a clear
\s-process signature in the isotopic composition of heavy trace elements.
These grains most likely condensed in the winds of a variety of AGB stars.
However, the non-solar and correlated Si and Ti isotopic compositions
measured in these grains are inconsistent with a pure \s-signature.
We present a possible solution to this much-discussed problem by
assuming a spread in the original composition of parent AGB stars due to
small chemical inhomogeneities in the interstellar medium at the time
of their birth. These inhomogeneities may naturally arise from variations
in the contributions of each nuclide to the interstellar medium by the
relevant stellar nucleosynthetic sites, SNII and different subtypes of
SNIa.
\end{abstract}
\section{AGB envelope prediction for Si and Ti}
Predictions from the AGB models discussed in \cite{g98,s97} for the Si
isotopic compositions in the envelope of AGB stars of solar metallicity
and different initial mass are shown in Fig. 1. They result from a
mixture of \s-processed material from the He shell and the envelope as a
consequence of recurrent third dredge-ups \cite{g98}, and are given
for six different choices of the amount of \ctb in the He intershell
(case ST refers to Fig. 1 of reference \cite{g98}).
Open symbols are for envelopes with
C/O$>$1, the condition for SiC condensation. Whereas predictions from
the same models match the measured compositions of heavy trace elements
in bulk SiC grains (Kr, Sr, Xe, Ba, Nd and Sm) \cite{g97}, and in
single SiC grains (Zr and Mo) \cite{gp98}, they cannot account for
the Si and Ti isotopic compositions of mainstream SiC grains
\cite{h94,h96}. As for Si, the measured isotopic ratios show a
much larger spread than AGB envelope predictions. Indeed,
the \s-process occurring in the He shell of AGB stars only slightly
affects the Si isotopic abundance (as well as most other light
neutron poisons), owing to their low neutron capture cross sections.
The resulting composition in the dredged-up material essentially
depends on the marginal activation of the
$^{22}$Ne neutron source during thermal pulses.

Analogous considerations hold for the Ti isotopic composition:
$\delta$($^{46,47}$Ti/$^{48}$Ti) values of individual SiC grains range from
-50 to +150 and from -50 to +70, respectively \cite{h94}, whereas
AGB envelope predictions cover a much smaller range, of up to +50 and
+10. Note that $\delta$($^{46,47}$Ti/$^{48}$Ti) are linearly correlated
with $\delta$($^{29,30}$Si/$^{28}$Si) \cite{h94}. In contrast, the
neutron-rich isotopes $^{49,50}$Ti are more easily produced by neutron
capture. Their $\delta$-values measured in SiC reach +150 and +300
\cite{h94}, in agreement with \s-process predictions.

\begin{figure}
\centerline
%{\vbox{\psfig{figure=nicsifig1.eps,height=9cm,width=\textwidth}}}
{\vbox{\psfig{figure=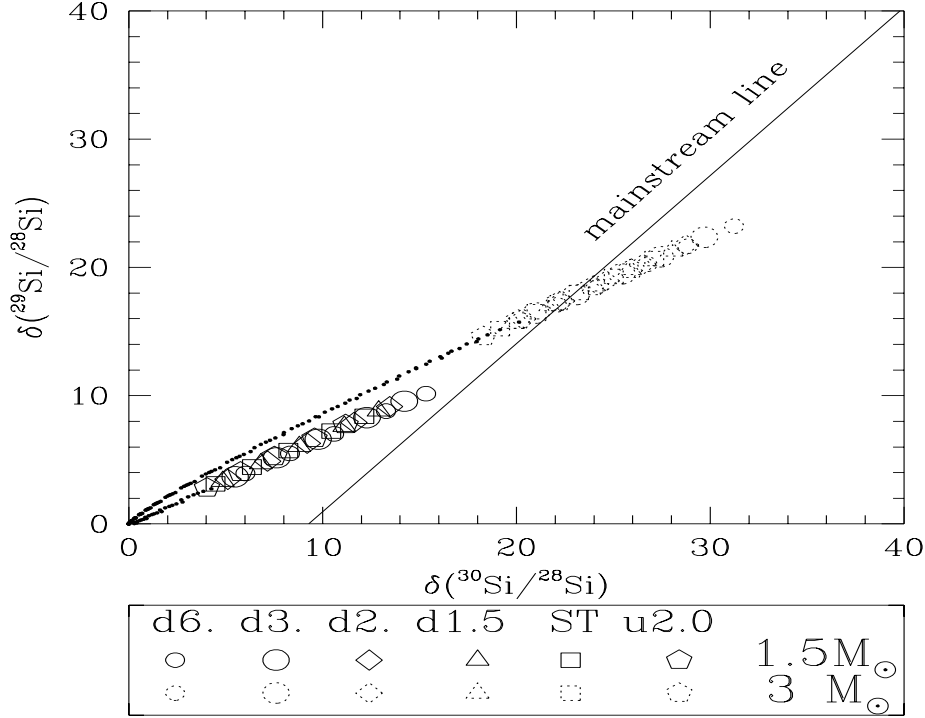,height=10 cm,width=\textwidth}}}
\caption[]{\small
\s-process model predictions for $\delta$($^{30}$Si/$^{28}$Si) and
$\delta$($^{29}$Si/$^{28}$Si) in the envelope of AGB stars of solar
metallicity and different initial mass, for six different
choices of the amount of \ctb supplied in the He intershell. The
$\delta$-values are permil deviations from the solar ratio:
e.g. $\delta$($^{30}$Si/$^{28}$Si)=
$[(^{30}$Si$/^{28}$Si$)/(^{30}$Si$/^{28}$Si$)_{\odot}-1]\times 1000$.
Open symbols are for C/O$>$1 in the
envelope. Note that the SiC mainstream isotopic compositions cover a much
larger range [from (-50,-100) to (+150,+250) permil in the Si three
isotope plot, see Fig. 2] than the predicted values, and the correlation
line has a slope of
$\sim$1.34.}
\end{figure}

\section {Galactic local inhomogeneities of Si and Ti}
A current interpretation of silicon isotopic ratios in SiC
relates them to the Galactic chemical evolution (GCE), taking into
account that the nucleosynthesis of $^{29,30}$Si is of secondary
nature, while $^{28}$Si is a primary isotope \cite{t96}. In this
view the mainstream correlation line reflects the evolution of
silicon isotopes with metallicity and leads to the conclusion that
the majority of presolar SiC grains should have been originated in
the outflows of AGB stars with higher than solar metallicity. To
overcome this contradiction, the hypothesis of a general
diffusion of stars from their birthplace towards higher
galactocentric distances has been advanced \cite{c97}.

Without discarding the GCE interpretation, we propose that the
interstellar medium (ISM) is affected by small local heterogeneities
and that this assumption may help to reach a better understanding of the Si
and Ti
isotopic
compositions observed in the grains.
According to
Woosley et al. \cite{w97}, three major stellar sources
in the Galaxy contribute
in different ways to the Si isotopes: (i) Supernovae
of Type {\rm II} (SN{\rm II}) \cite{w95}, (ii) Supernovae of 
Type {\rm I}a according to the standard model \cite{t86} and 
(iii) Supernovae of Type Ia originating from sub-Chandrasekhar 
white dwarfs accreting He from a binary companion \cite{w94}. 
As a test exercise, we started from an ISM of solar
composition and then perturbed it with small (positive and negative)
contributions by each of these three stellar sources.
In this way, the mainstream SiC grain parent stars, born in different
molecular clouds, are expected to show a spread of initial compositions,
reflecting
local heterogeneity in the ISM. We calculated the resulting perturbed
mass fractions in the ISM as
$X^i = X^i_{\odot} + \sum_{j=1}^{3} a_j M^i_j$, where $M^i_j$ is the
ejected mass (in solar masses) of isotope $i$ by the
stellar source $j$, and $a_j$ represents the level of contribution
from each source. A plausible range for the $a_j$ parameters has been
assumed, from $-$0.0004 $M^{-1}_{\odot}$ to $+$0.0004 $M^{-1}_{\odot}$.

\begin{figure}
\centerline
{\vbox{\psfig{figure=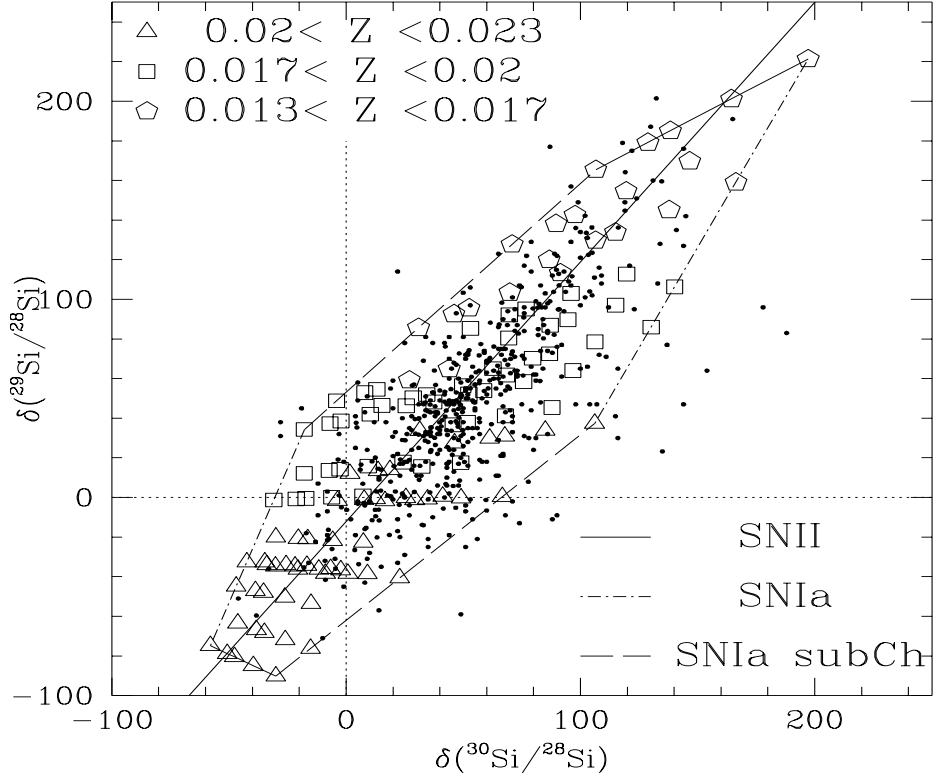,height=11.cm,width=\textwidth}}}
\caption[]{\small
Silicon $\delta$-values calculated from small inhomogeneities in the
ISM (open symbols). Different open symbols indicate slightly different
metallicities. The different kind of lines represent trends in the
variations due to the three different stellar sources of silicon
(see text for explanation).
Also shown as filled circles are the values measured in single SiC grains
\cite{h94,h96} and the mainstream correlation line.
}
\end{figure}
The resulting $\delta$-values are shown in Fig. 2 as open symbols and are
compared with the $\delta$-values measured in single SiC grains
(filled circles); the latter have errors (not shown in
the figure) ranging from 10 to 50 permil \cite{h94,h96}.
The mainstream correlation line is also shown in Fig. 2.
Since the heavy elements, in particular Fe, are produced in the
sources involved, the parent star metallicities will be perturbed
too: we obtain metallicities ranging from 0.013 to 0.023. This is
roughly in accord with the variation by a factor of two in [Fe/H]
for stars of the same age observed by Edvardsson et al. \cite{e93}.

Work is in progress in order to apply a similar model to
the Ti isotopic composition
of SiC grains. A preliminary solution has already been obtained
by including, beside the three major stellar sources mentioned above,
the rare type of SNIa described in \cite{wr97} and \cite{m98}, i. e.,
a massive white dwarf approaching the Chandrasekhar mass and
exploding while accreting mass in a binary system, which is likely
to contribute almost all galactic $^{50}$Ti.
Moreover, it has to be noted here that a complete astrophysical
interpretation of Si and Ti anomalies in SiC grains from AGB stars
must include also the small subpopulations of SiC
grains (up to a few percent of the total), such as SiC-Y and SiC-Z, which
show isotopic compositions different from those of the mainstream grains.

\acknowledgements{Maria Lugaro gratefully acknowledges the support of an
Overseas Postgraduates Research Scheme grant by the Australian Government
and of a travel grant by the Astronomical Society of Australia.}

\vfill
\end{document}